\UseRawInputEncoding

\documentclass[aps,prb,twocolumn,groupedaddress,showpacs, citeautoscript,superscriptaddress]{revtex4-1}
\usepackage{graphicx}% Include figure files
\usepackage{dcolumn}% Align table columns on decimal point
\usepackage{bm}% bold math
\usepackage{amsmath,amsfonts,amssymb,amsthm,verbatim}
\usepackage[ansinew]{inputenc}
\usepackage{color}

\usepackage{epsfig}
\usepackage{wasysym}

\begin{document}

\title{Magnetoelectric torque and edge currents in spin-orbit coupled graphene nanoribbons}

%\title{Bias voltage-free spin torque and edge currents in graphene nanoribbons caused by spin-orbit coupling}

%\title{Bias voltage-free spin torque, quantum spin Hall and quantum anomalous Hall effects in graphene nanoribbons caused by spin-orbit coupling}

%\title{Bias voltage-free spin torque, chiral charge and spin currents in graphene nanoribbons caused by spin-orbit coupling}

%\title{Gate-tunable spin torque and edge currents in magnetized graphene nanoribbons caused by spin-orbit coupling}

\author{Matheus S. M. de Sousa}

\affiliation{Department of Physics, PUC-Rio, 22451-900 Rio de Janeiro, Brazil}

\author{Manfred Sigrist}

\affiliation{Institute for Theoretical Physics, ETH Zurich, 8093 Zurich, Switzerland}

\author{Wei Chen}

\affiliation{Department of Physics, PUC-Rio, 22451-900 Rio de Janeiro, Brazil}

\date{\today}

\begin{abstract}

For graphene nanoribbons with Rashba spin-orbit coupling, the peculiar magnetic response due to the presence of a magnetization and geometric confinement are analyzed within a tight-binding model. We observe a sizable transverse susceptibility that can be considered as a gate voltage-induced magnetoelectric torque without the need of a bias voltage, with different directions for zigzag and armchair ribbons. The local torque generates non-collinear spin polarization between the two edges and/or along the ribbon, and the net torque averages to zero if the magnetization is homogeneous. Nevertheless, a nonzero net torque can appear in partially magnetized nanoribbons or in nanoflakes of irregular shapes. The equilibrium spin current produced by the spin-orbit coupling also appears in nanoribbons, but the component flowing in the direction of confinement is strongly suppressed. Even without the magnetization, an out-of-plane polarized chiral edge spin current is produced, resembling that in the quantum spin Hall effect. Moreover, a magnetization pointing perpendicular to the edge produces a laminar flow of edge charge currents, whose flow direction is symmetric (non chiral) or antisymmetric (chiral) between the two edges depends on whether the magnetization points in-plane or out-of-plane.

%These phenomena exist even in graphene nanoflakes that have open boundary in every direction.

\end{abstract}

\maketitle

\section{Introduction}

The celebrated Rashba spin-orbit coupling (RSOC) has a strong impact on the physical properties of two-dimensional (2D) metals with parabolic bands\cite{Bychkov84,Bychkov84_2}, especially their magnetic response and charge to spin interconversion. A number of these features originate from the RSOC induced spin-momentum locking, such as the Edelstein effect that causes a bias voltage-induced nonequilibrium in-plane spin polarization\cite{Edelstein90}.
In magnetized 2D systems, this effect can be exploited to induce magnetization dynamics known as the spin-orbit torque\cite{Manchon08,Manchon09,Haney10,Gambardella11,Pesin12,Haney13} , whose feasibility has been demonstrated extensively in experiments\cite{Miron10,Miron11,Manchon19}. In addition to these properties, RSOC also modifies the equilibrium properties of 2D metals, most notably causing an in-plane polarized spin current flowing throughout the system\cite{Rashba03,Sonin07,Tokatly08}.

Besides these effects occurring in infinitely extended 2D metals or electron gases (2DEG), RSOC also causes peculiar effects at the boundary of geometrically confined mesoscopic 2D systems. A notable example is the generation of out-of-plane polarized equilibrium edge spin currents\cite{Reynoso04,Grigoryan09}. Furthermore, in-plane spin polarization perpendicular to the edge can induce a persistent charge current decaying and oscillating in sign away from the edge\cite{Usaj05}. This behavior is very similar to that found in other spin-momentum locking systems in proximity to a magnet, such as for topological insulator/ferromagnetic metal junctions (TI/FMMs)\cite{Zegarra20,Chen20_TIFMM}. The tunability of RSOC by a gate voltage may provide a means to control such phenomena and engineer dedicated devices.

%{\cblue (1) How about QHE? When magnetization points out-of-plane it seems like I have to worry about QHE. Or say that we investigate the Zeeman splitting alone. QHE should include a gauge field in the hopping term, which is left for further investigation. Say many models did not consider QHE too. }

%Particularly in graphene, the recently uncovered gate-controllable SOC in graphene/transition metal dichalcogenide (TMD) heterostructures is fairly encouraging, since it points to electrically-controllable magnetic responses in these monolayer materials.

%{\cblue (1) Mention Uri20 which has measurement of persistent current on graphene due to topological magnetoelectric effect. They also claim to have a laminar edge current, but in a much larger scale, and presumably it is induced by the charge and topological magnetoelectric effect. }

In this paper, we use a tight-binding model to explore the spin torque and equilibrium currents caused by RSOC and geometric confinement in graphene nanoribbons. 
Numerous peculiarities make graphene nanoribbons a particularly intriguing system to study: (1) the linear dispersion near the two Dirac points that causes a unique spin-momentum locking profile\cite{Novoselov05,Zhou06,Geim07,CastroNeto09,Geim09,DasSarma11}; (2) the emergence of zero energy edge states in zigzag terminated ribbons \cite{Fujita96,Nakada96,Brey06,Ezawa06,Peres06,Akhmerov08,Tao11,Ruffieux16}: (3) the opening of energy gap depending on the ribbon width for armchair ribbons \cite{Son06,Son06_2,Barone06,Yang07,Han07,Kyle09}, and (4) the pronounced magnetic response at some sample edges \cite{Wakabayashi99,Lee05,Pisani07}. Moreover, it is important to note that an enhanced and gate-tunable spin-orbit coupling (SOC) has been demonstrated in graphene/transition-metal dichalcogenide (TMD) heterostructures\cite{Yang16,Wang16_3,Yang17,Safeer19,Ghiasi19,Benitez20}. In graphene/
yttrium iron garnet (YIG)\cite{Mendes15,Wang15_3,Dushenko16,Leutenantsmeyer17} and graphene/Co\cite{Rybkin18} heterostructures, both SOC and ferromagnetism are induced, which might be used as the stage for the phenomena we are going to discuss. Finally, the RSOC and magnetization are also known to cause quantum spin Hall (QSHE) and quantum anomalous Hall effects (QAHE) in graphene when combined with the intrinsic SOC\cite{Diniz13,Guassi15}.

We first elaborate on the feature that, in contrast to the usual longitudinal in-plane susceptibility in an infinitely large graphene, a transverse in-plane susceptibility develops in nanoribbons due to the RSOC and geometric confinement. The controllability of the RSOC suggests that this transverse susceptibility serves as a gate-voltage-induced magnetoelectric torque on the magnetization without the need of a bias voltage, contrary to the current-induced spin-orbit torque\cite{Dyrdal15,Li16,RodriguezVega17,Rybkin18,Zollner20,Rybkina20}. While this torque averages to zero for a homogeneously spin polarized system, well designed local variations in the magnetization can yield a net torque signal available to practical purpose. 
We further investigate the pattern of equilibrium spin currents in nanoribbons, especially concerning the symmetry of flow directions between the two edges, e.g. distinguished chiral and non-chiral structure. Charge currents induced by spin polarization exist in nanoribbons too, whose variability upon the direction of the magnetization will be a point of attention.

%This current has been linked to the asymmetric band structure due to distorted spin-momentum locking profile\cite{Zegarra20,Chen20_TIFMM}. However, we uncover that whether the flow direction is symmetric or antisymmetric between the two edges depends on the magnetization lies in-plane or out-of-plane, and so does the asymmetry of band structure. In contrast, the symmetry of the equilibrium spin current is unaltered by the presence of magnetization.

%in contrast to the edge currents in Chern insulators and quantum Hall effect (QHE) that flow in opposite direction at the two edges.

We start by introducing the lattice model for graphene with RSOC and spin magnetization, and then calculate the persistent spin current to show how the spin-momentum locking is modified by the presence of the magnetization in Sec.~\ref{sec:graphene}.
In Sec.~\ref{sec:nanoribbon}, we use a two-site toy model to demonstrate analytically the existence of a transverse susceptibility due to geometric confinement, which is then transferred to the discussion of both zigzag and armchair ribbons. In addition, we characterize the patterns and chiralities of persistent charge and spin currents, and also use an specific example to demonstrate that these phenomena also survive in graphene nanoflakes. Section \ref{sec:conclusions} summarizes our results.

\section{Extended tight-binding model of graphene \label{sec:graphene}}

%\subsection{Lattice model \label{sec:lattice_model}}

\begin{figure}[ht]
\begin{center}
\includegraphics[clip=true,width=0.99\columnwidth]{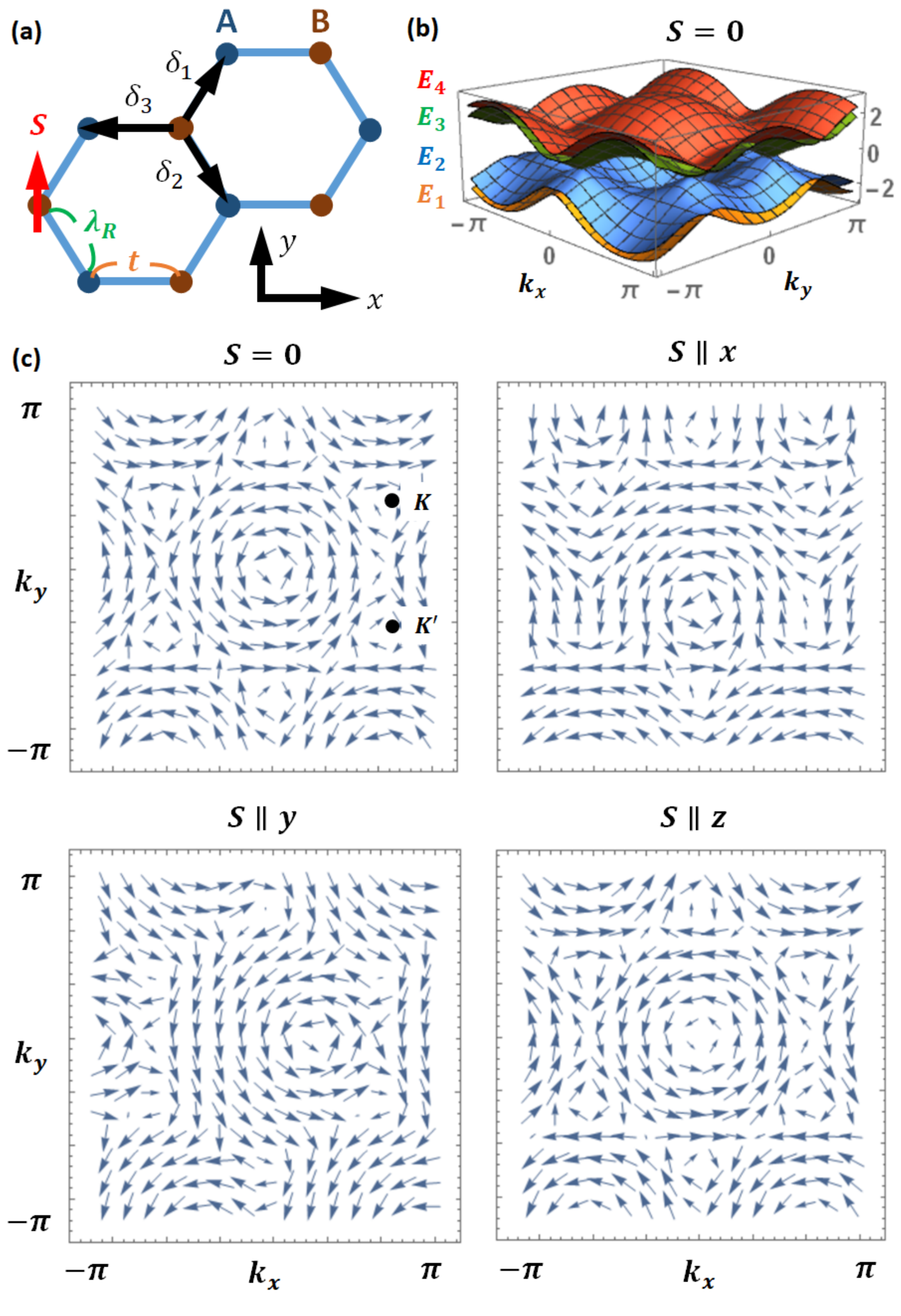}
\caption{(a) The definition of coordinate in our honeycomb lattice. (b) The resulting band structure in the presence of RSOC without magnetization ${\bf S}={\bf 0}$. (c) The spin-momentum locking profile of the lowest eigenstate $|u_{{\bf k}1}\rangle$ in the absence ${\bf S}={\bf 0}$ and presence ${\bf S}\parallel\left\{{\hat{\bf x}},{\hat{\bf y}},{\hat{\bf z}}\right\}$ of the magnetization. }
\label{fig:honeycomb_lattice}
\end{center}
\end{figure}

In order to emphasize the impact of geometric confinement, we start by addressing the spintronics related effects due to RSOC for an infinite graphene sheet, before discussing the same physics for graphene ribbons and nanoflakes. 
As indicated in Fig.~\ref{fig:honeycomb_lattice} (a), three lattice vectors (in units of bond length $a=1$) characterize the honeycomb lattice,
\begin{eqnarray}
{\boldsymbol\delta}_{1}=\left(\frac{1}{2},\frac{\sqrt{3}}{2}\right),\;\;\;
{\boldsymbol\delta}_{2}=\left(\frac{1}{2},-\frac{\sqrt{3}}{2}\right),\;\;\;
{\boldsymbol\delta}_{3}=\left(-1,0\right).\;\;\;
\label{delta_definition}
\end{eqnarray}
connecting neighbouring lattice sites belonging to the two different sublattices, $A$ and $B$. 
We now formulate the following tight-binding model that incorporates both RSOC and spin polarization\cite{Kane05_2,Kane05,Min06,Dyrdal17,Peralta19}
\begin{eqnarray}
H&=&-t\sum_{\langle ij\rangle,\sigma}c_{i\sigma}^{\dag}c_{j\sigma}+J_{ex}\sum_{i,\alpha,\beta}{\bf S}\cdot c_{i\alpha}^{\dag}{\boldsymbol \sigma}_{\alpha\beta}c_{i\beta}
\nonumber \\
&+&i\lambda_{R}\sum_{\langle ij\rangle,\alpha,\beta}c_{i\alpha}^{\dag}\left({\boldsymbol \sigma}_{\alpha\beta}\times{\bf d}_{ij}\right)^{z}c_{j\beta}-\mu\sum_{i\sigma}c_{i,\sigma}^{\dag}c_{i\sigma}.
\label{Hamiltonian_graphene_Rashba_mag}
\end{eqnarray}
Here $ c^{\dag}_{i \sigma} (c_{i \sigma}) $ creates (annihilates) an electron of spin $ \sigma $ on the lattice site $ i$, where sum with $ \langle i j \rangle $ run over nearest neighbor lattice sites. The hopping matrix element is $ - t $, $\lambda_{R}$ is the RSOC coupling constant, ${\boldsymbol \sigma}=(\sigma^{x},\sigma^{y},\sigma^{z})$ are the spin Pauli matrices, ${\bf d}_{ij}$ is the vector connecting the site $i$ to $j$, $J_{ex}$ is the exchange coupling between the magnetization ${\bf S}=S(\sin\theta\cos\phi,\sin\theta\sin\phi,\cos\theta)$ and the spin, and $\mu$ is the chemical potential. The bipartite structure of our system allows us to use a sublattice formulation with $I=\left\{A,B\right\}$, where we define the basis $\psi=\left(A\uparrow,B\uparrow,A\downarrow,B\downarrow\right)$, and the electron operators in Eq.~(\ref{Hamiltonian_graphene_Rashba_mag}) can be split into $c_{i\sigma}\rightarrow\left\{c_{Ai\sigma},c_{Bi\sigma}\right\}$, where $i$ now denotes the position of each unit cell containing an $A$ and $B$ lattice site. After Fourier transformation $c_{Ii\sigma}=\sum_{\bf k}e^{i{\bf k\cdot r}_{i}}c_{I{\bf k}\sigma}$, the Hamiltonian $H=\sum_{\bf k IJ\alpha\beta}c_{I{\bf k}\alpha}^{\dag}H_{I\alpha J\beta}({\bf k})c_{J{\bf k}\beta}$ defines the matrix
\begin{eqnarray}
&&H_{I\alpha J\beta}({\bf k})=
\nonumber \\
&&\left(
\begin{array}{cccc}
J_{ex}S^{z} & tZ^{\ast} & J_{ex}S_{\perp}e^{-i\phi} & \lambda_{R}Y^{\ast} \\
tZ & J_{ex}S^{z} & \lambda_{R}X^{\ast} & J_{ex}S_{\perp}e^{-i\phi} \\
J_{ex}S_{\perp}e^{i\phi} & \lambda_{R}X & -J_{ex}S^{z} & tZ^{\ast} \\
\lambda_{R}Y & J_{ex}S_{\perp}e^{i\phi} & tZ & -J_{ex}S^{z}
\end{array}
\right),\;\;\;\;
\label{Hamiltonian_matrix_form}
\end{eqnarray}
with 
\begin{eqnarray}
&&Z\equiv e_{1}+e_{2}+e_{3},
\nonumber \\
&&X\equiv\frac{-1-i\sqrt{3}}{2}e_{1}^{\ast}+\frac{-1+i\sqrt{3}}{2}e_{2}^{\ast}+e_{3}^{\ast},
\nonumber \\
&&Y\equiv\frac{1+i\sqrt{3}}{2}e_{1}+\frac{1-i\sqrt{3}}{2}e_{2}-e_{3},
\nonumber \\
\end{eqnarray}
where we have defined $e_{a}=e^{i{\bf k}\cdot{\boldsymbol\delta}_{a}}$ and $S_{\perp}=\sqrt{(S^{x})^{2}+(S^{y})^{2}}=S\sin\theta$ ($ \theta \in [0,\pi]$). In the following we choose the parameters
\begin{eqnarray}
&&J_{ex}=0.2,\;\;\lambda_{R}=0.2,
\;\;\mu=0.5,\;\;k_{B}T=0.03,\;\;\;\;
\label{graphene_parameters}
\end{eqnarray}
with $ t=1 $ as energy unit, 
such that the magnetic response is more pronounced, but we emphasize that their patterns are fairly robust against variation of these parameters.

Diagonalizing the Hamiltonian yields two filled bands $E_{{\bf k}1}<E_{{\bf k}2}<0$ and two empty bands $0<E_{{\bf k}3}<E_{{\bf k}4}$, with the corresponding four eigenstates $\left\{|u_{{\bf k}1}\rangle,|u_{{\bf k}2}\rangle,|u_{{\bf k}3}\rangle,|u_{{\bf k}4}\rangle\right\}$. These four bands are shown in Fig.~\ref{fig:honeycomb_lattice} (b) for zero magnetization. In this case the spin expectation value $\langle{\boldsymbol \sigma}\rangle_{{\bf k}\eta}=\langle u_{{\bf k}\eta}|{\boldsymbol \sigma}|u_{{\bf k}\eta}\rangle$ for each eigenstate has only inplane components $\left\{\langle \sigma^{x}\rangle_{{\bf k}\eta},\langle \sigma^{y}\rangle_{{\bf k}\eta}\right\}$, as displayed in the vector plots in Fig.~\ref{fig:honeycomb_lattice} (c) for $ S=0 $, where we also see $\langle{\boldsymbol \sigma}\rangle_{{\bf k}1}=-\langle{\boldsymbol \sigma}\rangle_{{\bf k}2}$. We see that the spin texture forms a vortex around each Dirac point, ${\bf K}=(2\pi/3,2\pi/3\sqrt{3})$ and ${\bf K'}=(2\pi/3,-2\pi/3\sqrt{3})$ with the same vorticity, but opposite vorticity appears for the vortex around the origin ${\bf k}=(0,0)$. This spin pattern represents the spin-momentum locking due to the RSOC.

%{\cblue (1) I have checked that in the absence of RSOC, the two filled bands $\langle{\boldsymbol \sigma}\rangle_{{\bf k}1}=-\langle{\boldsymbol \sigma}\rangle_{{\bf k}2}$ have exactly opposite spin expectation value, and for each of them the vorticity of spin momentum locking around the ${\bf K}$ and ${\bf K}'$ points are the same.  }

%\begin{figure}[ht]
%\begin{center}
%\includegraphics[clip=true,width=0.9\columnwidth]{graphene_Rashba_result}
%\caption{Graphene with Rashba interaction $\lambda_{R}=0.2$ but no magnetization. }
%\label{fig:graphene_Rashba_result}
%\end{center}
%\end{figure}

As common in 2D Rashba metals, the presence of a magnetization ${\bf S}$ is expected to distort the profile of spin pattern \cite{Manchon09,Gambardella11}, as demonstrated for magnetizations along the three principle directions ${\bf S}\parallel\left\{{\hat{\bf x}},{\hat{\bf y}},{\hat{\bf z}}\right\}$ in Fig.~\ref{fig:honeycomb_lattice} (c), which shows the planar components of $\langle{\boldsymbol \sigma}\rangle_{{\bf k}1}$ only. For ${\bf S}\parallel{\hat{\bf x}}$, the $\langle{\boldsymbol \sigma}\rangle_{{\bf k}1}$ becomes asymmetric between $+k_{y}$ and $-k_{y}$, which also causes an asymmetry in the dispersion in this direction. Analogously, ${\bf S}\parallel{\hat{\bf y}}$, yields an asymmetry between $+k_{x}$ and $-k_{x}$, whereas for ${\bf S}\parallel{\hat{\bf z}}$, the planar components basically remain the same as that in the absence of magnetization. By calculating the spin polarization in our lattice model with periodic boundary condition (PBC), we also verify that the spin polarization on either sublattice is always longitudinal, i.e., along the direction of magnetization.

%in a large enough graphene $\apprge 10\times 10$ with periodic boundary condition (PBC) in both ${\hat{\bf x}}$ and ${\hat{\bf y}}$ directions.

%From the pattern of the distorted spin-momentum locking, it is also clear that at zero temperature, the spin polarization per site is always along the direction of magnetization ${\bf S}$. For instance, for the ${\bf S}\parallel{\hat{\bf x}}$ panel in Fig.~\ref{fig:honeycomb_lattice} (c), the spin polarization at ${\bf k}$ and $-{\bf k}$ satisfy $\langle s^{y}\rangle_{{\bf k}1}=-\langle s^{y}\rangle_{-{\bf k}1}$, so they cancel after a summation.

%{\cblue (1) Now calculate the spin expectation value on every site from the eigenstates of Eq.~(\ref{Hamiltonian_matrix_form}) and verify whether the spin polarization is always along the direction of magnetization.  }

%\begin{figure}[ht]
%\begin{center}
%\includegraphics[clip=true,width=0.9\columnwidth]{graphene_Rashba_Salongx_result}
%\caption{Graphene with Rashba interaction $\lambda_{R}=0.2$ and $J_{ex}S=0.2$ at magnetization along ${\bf S}\parallel{\hat{\bf x}}$ (equivalently $\theta=\pi/2$ and $\phi=0$). }
%\label{fig:graphene_Rashba_Salongx_result}
%\end{center}
%\end{figure}

The charge and spin current density operators are constructed from the equations of motion of the charge and spin density operator, respectively, which correspond to continuity equations\cite{Zegarra20,Chen20_TIFMM}
\begin{eqnarray}
\dot{n}_{i}&=&\frac{i}{\hbar}\left[H,n_{i}\right]=\frac{i}{\hbar}\left[H_{t}+H_{R},n_{i}\right]
\nonumber \\
&=&-{\boldsymbol\nabla}\cdot{\bf J}_{i}^{0}=-\frac{1}{a}\sum_{\eta}J_{i,i+\eta}^{0},
\nonumber \\
\dot{m}_{i}^{a}&=&\frac{i}{\hbar}\left[H,m_{i}^{a}\right]=\frac{i}{\hbar}\left[H_{t}+H_{R},m_{i}^{a}\right]
+\frac{i}{\hbar}\left[H_{J},m_{i}^{a}\right]
\nonumber \\
&=&-{\boldsymbol\nabla}\cdot{\bf J}_{i}^{a}+\tau_{i}^{a}
\nonumber \\
&=&-\frac{1}{a}\sum_{\eta}J_{i,i+\eta}^{a}+\frac{2J_{ex}}{\hbar}\left({\bf S}\times c_{i\alpha}{\boldsymbol\sigma}_{\alpha\beta}c_{i\beta}\right)^{a},
\label{charge_spin_continuity_eq}
\end{eqnarray}
where ${\boldsymbol\eta}=\left(-{\boldsymbol\delta_{1}},-{\boldsymbol\delta_{2}},-{\boldsymbol\delta_{3}}\right)$ if $i\in A$, and ${\boldsymbol\eta}=\left({\boldsymbol\delta_{1}},{\boldsymbol\delta_{2}},{\boldsymbol\delta_{3}}\right)$ if $i\in B$. Moreover, $\tau_{i}^{a}=\frac{2J_{ex}}{\hbar}\left({\bf S}\times c_{i\alpha}{\boldsymbol\sigma}_{\alpha\beta}c_{i\beta}\right)^{a}$ denotes the spin torque term as originating from the usual Landau-Lifshitz dynamics. The calculation of the commutators in Eq.~(\ref{charge_spin_continuity_eq}) is detailed in Appendix \ref{apx:charge_spin_current_operator}.

\section{Graphene nanoribbons with magnetization and Rashba spin-orbit coupling \label{sec:nanoribbon}}

\subsection{Two-site toy model \label{sec:two_site_toy_model}}

For a deeper insight into the impact of geometrical confinement on properties of graphene with RSOC and magnetization, we first present an exactly solvable two-site toy model to demonstrate the feature of the transverse spin susceptibility and the persistent spin current. Similar models have been proposed to explain the microscopic mechanism of noncollinear magnetic order and equilibrium spin currents\cite{Katsura05,Bruno05,Chen14_two_coupled_FM}, while we will put emphasis on the spin torque here. The two-site model Hamiltonian reads, 
\begin{eqnarray}
H&=&\sum_{\sigma}t\left(c_{A\sigma}^{\dag}c_{B\sigma}+c_{B\sigma}^{\dag}c_{A\sigma}\right)
+J_{ex}\sum_{i=A,B}{\bf S}_{i}\cdot c_{i\alpha}^{\dag}{\boldsymbol \sigma}_{\alpha\beta}c_{i\beta}
\nonumber \\
&+&i\lambda_{R}\sum_{ij}c_{i\alpha}^{\dag}\left({\boldsymbol \sigma}_{\alpha\beta}\times{\bf d}_{ij}\right)^{z}c_{j\beta},
\label{Hamiltonian_two_site_model}
\end{eqnarray}
where the two sites $i=\left\{A,B\right\}$ are assumed to be connected along ${\bf d}_{AB}=-{\bf d}_{BA}\parallel{\hat {\bf x}}$, and we consider the same magnetization on the two sites ${\bf S}_{A}={\bf S}_{B}$. The two sites feature the $A$ and $B$ sublattice sites in the graphene unit cell connected along ${\hat{\bf x}}$ direction, as implied in Fig.~\ref{fig:honeycomb_lattice}. The $4\times 4$ Hamiltonian in the basis of $(c_{A\uparrow},c_{B\uparrow},c_{A\downarrow},c_{B\downarrow})$ is
\begin{eqnarray}
H=\left(\begin{array}{cccc}
J_{ex}S^{z} & t & J_{ex}S_{\perp}e^{-i\phi} & -\lambda_{R} \\
t & J_{ex}S^{z} & \lambda_{R} & J_{ex}S_{\perp}e^{-i\phi} \\
J_{ex}S_{\perp}e^{i\phi} & \lambda_{R} & -J_{ex}S^{z} & t \\
-\lambda_{R} & J_{ex}S_{\perp}e^{i\phi} & t & -J_{ex}S^{z}
\end{array}\right),
\nonumber \\
\end{eqnarray}
analogous to Eq.(\ref{Hamiltonian_matrix_form}). 
Assuming $\left\{J_{ex},\lambda_{R},t\right\}>0$ and $\left\{J_{ex},\lambda_{R}\right\}\ll t$, for any magnetization direction there are two negative ($E_1,E_2$) and two positive ($E_3,E_4$) eigenenergies. Suppose the two negative eigenstates $|u_{1}\rangle$ and $|u_{2}\rangle$ are occupied. Then the $\alpha$ component spin expectation value on site $i$ is given by $\langle \sigma_{i}^{\alpha}\rangle=\langle u_{1}|\sigma_{i}^{\alpha}|u_{1}\rangle+\langle u_{2}|\sigma_{i}^{\alpha}|u_{2}\rangle$, which can be expanded in powers of $J_{ex}S$ to obtain the susceptibility along principle directions $\chi_{i}^{\alpha\beta}=\partial \langle \sigma_{i}^{\alpha}\rangle/\partial (J_{ex}S^{\beta})$. We find 
\begin{eqnarray}
&&\chi_{A}^{\alpha\beta}=\left(\begin{array}{ccc}
\chi_{A}^{xx} & \chi_{A}^{xy} & \chi_{A}^{xz} \\
\chi_{A}^{yx} & \chi_{A}^{yy} & \chi_{A}^{yz} \\
\chi_{A}^{zx} & \chi_{A}^{zy} & \chi_{A}^{zz}
\end{array}\right)
\nonumber \\
&&=\frac{1}{(t^{2}+\lambda_{R}^{2})^{3/2}}\left(\begin{array}{ccc}
-\lambda_{R}^{2} & 0 & t\lambda_{R} \\
0 & 0 & 0 \\
-t\lambda_{R} & 0 & -\lambda_{R}^{2}
\end{array}\right)=\left(\chi_{B}^{\alpha\beta}\right)^{T}.
\label{chi_tensor_2site}
\end{eqnarray}
Thus there is a transverse response $\left\{\chi^{xz},\chi^{zx}\right\}$ between $x$ and $z$ directions, and consequently a spin torque on both sites due to Landau-Lifshitz dynamics,
\begin{eqnarray}
\frac{d{\bf S}_{i}}{dt}=\frac{J_{ex}}{\hbar}\langle{\boldsymbol\sigma}_{i}\rangle\times{\bf S}_{i}.
\label{Landau_Lifshitz_dynamics}
\end{eqnarray}\
However, because the transverse susceptibility is opposite on the two sites $\chi_{A}^{xz}=-\chi_{B}^{xz}$, the net torque $\sum_{i=A,B}d{\bf S}_{i}/dt$ vanishes. If we assume a coupling of the two magnetic moments by some exchange interaction $J_{AB}{\bf S}_{A}\cdot{\bf S}_{B}$, then the transverse spin polarization causes a canting angle $\theta_{AB}\approx 2\cos^{-1}\left(J_{ex}^{2}\lambda_{R}/2J_{AB}t^{2}\right)$ between them in the $xy$-plane, realizing a Dzyaloshinskii–-Moriya interaction (DMI)\cite{Dzyaloshinsky58,Moriya60}. Note that $\chi_{A}^{xx}$ and $\chi_{A}^{zz}$ in Eq.~(\ref{chi_tensor_2site}) represent the corrections to the longitudinal spin polarization due to RSOC, which must be negative, since the spins would be fully polarized $\langle {\boldsymbol \sigma}_{i}\rangle=1$ along $ {\bf S} $ if RSOC were absent. The $\chi_{A}^{yy}=0$ means the magnetic response in the ${\hat{\bf y}}$ direction is unaffected.

%{\cblue (1) Think about how to refer to my TI Edelstein effect\cite{Chen20_TI_Edelstein} and SHE-STT\cite{Chen15_spin_transfer_torque} papers. }

%Besides, in the absence of the RSOC $\lambda_{R}=0$, both sites are fully spin polarized $\langle s_{i}^{\alpha}\rangle=1$ along the magnetization direction ${\bf S}\parallel{\hat{\boldsymbol\alpha}}$ at arbitrarily small $J_{ex}$.

The local charge and spin current operators in this toy model can be constructed from the same formalism used in Sec.~\ref{sec:graphene}. Denoting the current operator $J_{AB}^{a}$ as the one flowing from site $A$ to $B$ and $J_{BA}^{a}$ for the opposite direction, we find that for either ${\bf S}\parallel{\hat{\bf x}}$ or ${\bf S}\parallel{\hat{\bf z}}$, the only nonzero expectation value is the spin current $\langle J^{y}\rangle$, and it satisfies
\begin{eqnarray}
&&\langle J_{AB}^{y}\rangle=-\langle J_{BA}^{y}\rangle
\nonumber \\
&&=\frac{J_{ex}S\lambda_{R}}{\sqrt{(t-J_{ex}S)^{2}+\lambda_{R}^{2}}}
-\frac{J_{ex}S\lambda_{R}}{\sqrt{(t+J_{ex}S)^{2}+\lambda_{R}^{2}}}.
\end{eqnarray}
There is no spin current for ${\bf S}\parallel{\hat{\bf y}}$. For all cases the charge current is always absent, but all three components of the spin current can be non-zereo, in general. Moreover, the $x$ and $z$ components of the spin currents flowing in the two directions are not negative of each other $\langle J_{AB}^{x,z}\rangle\neq -\langle J_{BA}^{x,z}\rangle$, but the continuity equation $\langle\dot{m}_{A}^{a}\rangle=\langle\dot{m}_{B}^{a}\rangle=0$ is explicitly satisfied, if the spin torque is taken into account as in Eq.~(\ref{charge_spin_continuity_eq}).

%Because the model in Eq.~(\ref{Hamiltonian_two_site_model}) does not cause a net torque, we proceed to examine a modified version where only site $A$ has magnetization, i.e., the exchange term contains only $J_{ex}{\bf S}_{A}\cdot c_{A\alpha}^{\dag}{\boldsymbol \sigma}_{\alpha\beta}c_{A\beta}$. The same calculation yields the susceptibility on site $A$
%\begin{eqnarray}
%\chi_{A}^{\alpha\beta}=....
%\end{eqnarray}
%indicating a spin torque on ${\bf S}_{A}$. This result suggests that to cultivate this equilibrium spin torque, one must engineer a situation that only part of the system has magnetization. We shall see a similar result in the following sections.

%{\cblue (1) I have checked that in if in the two-site model only the $A$ site has magnetization, then the susceptibility on the $A$ site has only longitudinal but no transverse components, so there is no torque. So this 2-site-1-magnetization model is not really supportive.  }

\begin{figure*}[htpb]
\includegraphics[width=1.6\columnwidth]{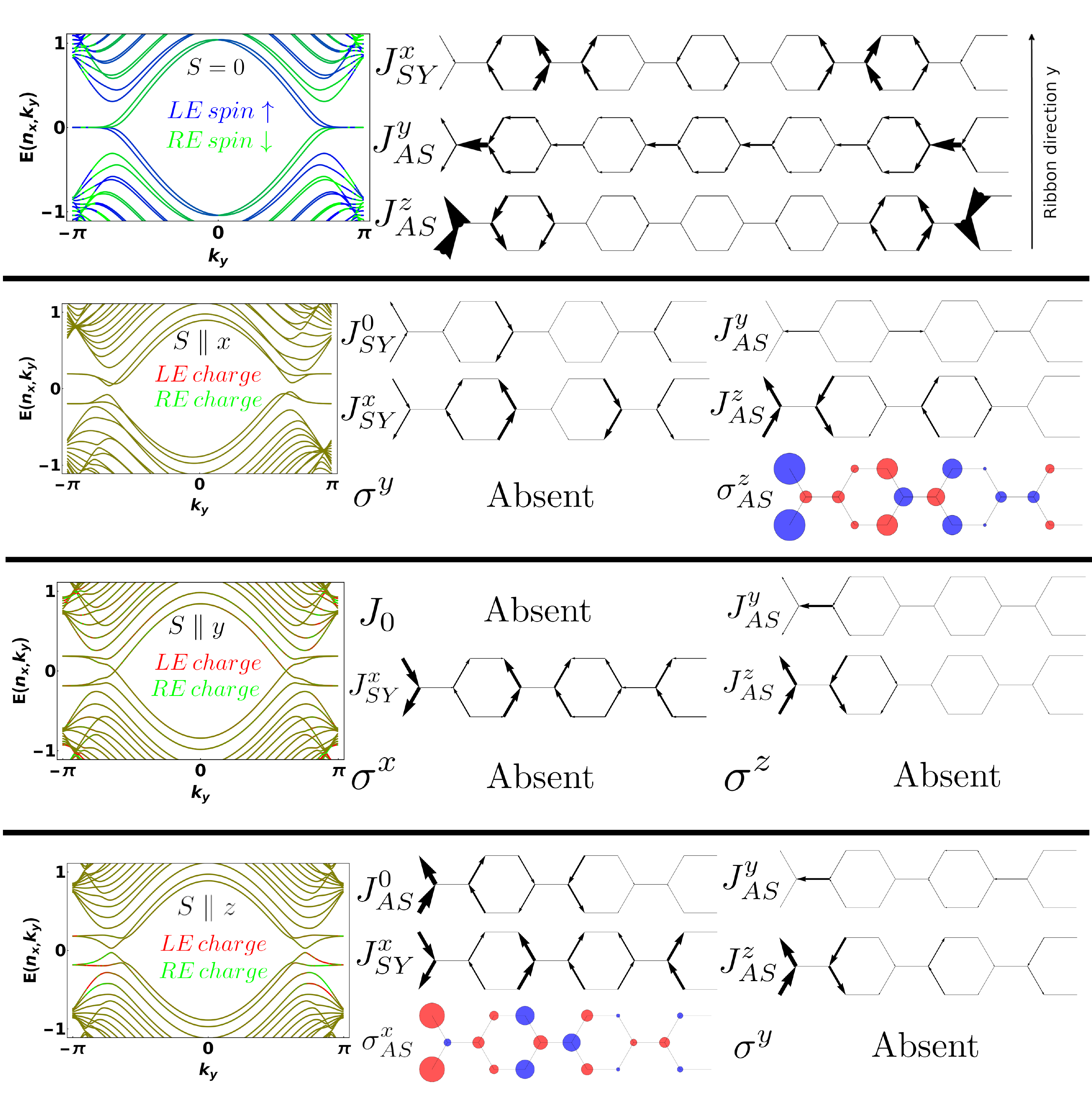}
\caption{The band structure $E(n_{x},k_{y})$, persistent charge and spin currents $J^{\alpha}$, and transverse spin polarization $\sigma^{\alpha}$ (red positive and blue negative) in the mirror symmetric zigzag ribbon with RSOC at different magnetization directions ${\bf S}$. The pattern of these quantities repeats along the ribbon direction ${\hat{\bf y}}$, and is either symmetric (labeled SY) or antisymmetric (labeled AS) under $P_{\rm zig}^{\rm mir}$ in Eq.~(\ref{zigzag_symmetry_operator}), so we only present the left half of a unit cell, except for ${\bf S}={\bf 0}$ we show the entire unit cell. For the absence of magnetization case ${\bf S}={\bf 0}$, the colors in the band structure indicate the spin up and down components at the left half space (LE) for the eigenstates $|n_{x},k_{y}\rangle$, and for ${\bf S}\neq{\bf 0}$ cases they indicate the probability of the eigenstates at the left and right (RE) half space. The largest arrow in the figure has magnitude $0.015$, and the largest disk $0.011$.  }
\label{fig:zigzag_ribbon_current}
\end{figure*}

\subsection{Persistent currents and spin torques in Zigzag ribbons \label{sec:persistent_current_zigzag}}

The results in the two-site model suggests the existence of spin torque and equilibrium spin current in geometrically confined Rashba systems, which motivates us to investigate our graphene model in Sec.~\ref{sec:graphene} in the nanoribbon geometry. We first consider the zigzag ribbon by taking open boundary condition (OBC) in the ${\hat{\bf x}}$ direction and PBC along ${\hat{\bf y}}$ direction for our lattice model in Eq.~(\ref{Hamiltonian_graphene_Rashba_mag}). There are two kinds of zigzag ribbons as far as the symmetry between the two edges is concerned, namely the mirror symmetric and the glide plane symmetric ones, each invariant under the corresponding symmetry operation
\begin{eqnarray}
P_{\rm zig}^{\rm mir}=\left\{P_{x}|\,{\bf 0}\right\},\;\;\;
P_{\rm zig}^{\rm gli}=\left\{P_{x}|\,\sqrt{3}/2{\hat{\bf y}}\right\},
\label{zigzag_symmetry_operator}
\end{eqnarray}
using the common notation of space group operation, where $P_{x}$ denotes the mirror reflection in ${\hat{\bf x}}$ directions with respect to the central axis of the ribbon. We find that the symmetries of the patterns of the currents $J^{\alpha}$ and spin polarization $\sigma^{\alpha}$ in the mirror symmetric ribbon defined with respect to $P_{\rm zig}^{\rm mir}$ are the same as those in the glide plane symmetric ribbon defined with respect to $P_{\rm zig}^{\rm gli}$, so we only present the former ones for simplicity. In particular, we use the zigzag ribbon of 24-site width as an example, as shown in Fig.~\ref{fig:zigzag_ribbon_current}. The patterns are translationally invariant along the ribbon direction ${\hat{\bf y}}$, and are either symmetric (labeled SY) or antisymmetric (labeled AS) under $P_{\rm zig}^{\rm mir}$, such that we only show the left half of a unit cell, with size of the arrows and disks indicating the magnitude of the currents and spin polarization, respectively. In addition, to clarify the origin of the edge charge current, for each eigenstate $|n_{x},k_{y}\rangle$ we calculate the weight of the wave function closer to the left (LE) and right (RE) edges
\begin{eqnarray}
&&n_{n_{x},k_{y}}^{LE}=\sum_{1\leq x\leq N_{x}/2}|\psi_{n_{x},k_{y}}(x)|^{2},
\nonumber \\
&&n_{n_{x},k_{y}}^{RE}=\sum_{N_{x}/2+1\leq x\leq N_{x}}|\psi_{n_{x},k_{y}}(x)|^{2}.
\label{zigzag_edge_charge}
\end{eqnarray}
Likewisely, to understand the spin current $J^{z}$ at the left edge, we calculate the spin polarization $\sigma^{z}$ at the left edge for each eigenstate
\begin{eqnarray}
m_{n_{x},k_{y}}^{z,LE}=\sum_{1\leq x\leq N_{x}/2}\langle\sigma^{z}_{n_{x},k_{y}}(x)\rangle.
\label{zigzag_edge_spin}
\end{eqnarray}
These quantities are represented by colors in the band structure. The results are summarized below according to different magnetization directions.

(i) ${\bf S}={\bf 0}$ : In the absence of magnetization, the $\left\{J^{x},J^{y}\right\}$ flowing along $\left\{{\hat{\bf y}},{\hat{\bf x}}\right\}$ induced by RSOC are present in the zigzag ribbon, whereby $J^{y}$ is largely suppressed due to OBC in ${\hat{\bf x}}$\cite{Sun07}. A helical (spin chiral) edge spin current $J^{z}$ is also produced, as hinted by the results in 2DEG\cite{Reynoso04,Grigoryan09,Nakhmedov12}, although in zigzag ribbons it exists even without a magnetic field and demonstrates helicity. In Fig.\ref{fig:zigzag_ribbon_current} the colored band structure clarifies the origin of $J^{z}$: for every left edge (LE) spin up $k_{y}$ state (blue) there exists a corresponding LE spin down $-k_{y}$ state (green), yielding counter propagating spins at the LE. This is true for all eigenstates, so the finite $J^{z}$ is not only the result of low energy states, similar to that has been discussed recently for the QSHE in topological insulators\cite{Chen20_absence_edge_current}. These features for the spin currents remain true even in the presence of a finite magnetization ${\bf S}\neq{\bf 0}$.

(ii) ${\bf S}\parallel{\hat{\bf x}}$ : It is known that in 2DEG with RSOC, an in-plane magnetization pointing perpendicular to the edge produces an edge charge current, whose flow direction depends on the distance away from the edge\cite{Usaj05}. %although its chirality remains unclear. 
${\bf S}\parallel{\hat {\bf x}}$ corresponds to this situation, in which we indeed see a charge current $J^{0}$ that is symmetric (nonchiral) between the two edges. The eigenstates are not particularly localized at either edge (the band structure is not particularly red or green), but the band structure becomes asymmetric $E(n_{x},k_{y})=-E(n_{x},-k_{y})$, which can cause $J^{0}$ along the ribbon, a mechanism that has been pointed out for a superconductor/noncollinear magnet\cite{Chen15_Majorana} and TI/FMM heterostructures\cite{Zegarra20,Chen20_TIFMM}. A large transverse spin polarization $\sigma^{z}$ is induced near the edge without any bias voltage (in contrast to that produced by a bias voltage\cite{Nomura05,Bokes10,Erlingsson07,Zhang14_2,Khaetskii14,Khaetskii17,Cysne21}), signaling the existence of a local torque. The torque is antisymmetric between the two edges and hence averages to zero, and is expected to cause a non-collinear order between the two edges as in the two-spin model in Sec.~\ref{sec:two_site_toy_model}. For the parameters in Eq.~(\ref{graphene_parameters}), the largest edge currents are of the order of $\sim 0.01$, in units of $et/\hbar\sim 10^{-4}$A for $J_y^{0}$ and $\mu_{B}t/\hbar\sim 10^{15}\mu_{B}/$s for $\left\{J^{x},J^{y},J^{z}\right\}$, where $\mu_{B}$ is the Bohr magneton, and the largest edge spin polarizations are of the order of $\sim 0.01\mu_{B}$.

(iii) ${\bf S}\parallel{\hat{\bf y}}$ : For the case of a magnetization along the ribbon, we find no charge current $J^{0}$ and no transverse spin polarization $\sigma^{x}=\sigma^{z}=0$, and hence there is no local torque. The energy spectrum is half-metallic.

(iv) ${\bf S}\parallel{\hat{\bf z}}$ : Interestingly, we find that an out-of-plane magnetization also produces a charge current $J^{0}$, but it is antisymmetric (chiral) between the two edges. Comparing with that in the ${\bf S}\parallel{\hat{\bf x}}$ case, this suggests the chirality of $J^{0}$ can be controlled by the orientation of the magnetization. This current only occurs when the chemical potential is finite $\mu\neq 0$, similar to that in the QAHE in Chern insulators\cite{Chen20_absence_edge_current}, although our spectrum remains gapped. The band structure is symmetric between $+k_{y}$ and $-k_{y}$ at ${\bf S}\parallel{\hat {\bf z}}$, but the wave function distribution is not: If $+k_{y}$ state is mostly localized at the left edge (red), then the $-k_{y}$ state is more localized at the right edge (green), suggesting counter propagating charge currents at the two edges. Note that some of the low energy states are inherited from the flat band edge states of the pristine zigzag ribbon, which become dispersive and chiral under the influence of magnetization and RSOC. Finally, a transverse spin polarization $\sigma^{x}$ occurs near the edge, which are opposite at the two edges and hence average to zero.

%so an asymmetric band structure is in fact not a necessary condition for this laminar current. We conclude that the charge current occurs whenever the magnetization is perpendicular to the edge, not necessarily has to be in-plane. The net current is zero, suggesting that local measurements are required to detect this edge current, which may be quite challenging experimentally.

%Finally, for the case of magnetization along the ribbon ${\bf S}\parallel{\hat {\bf y}}$ or no magnetization ${\bf S}={\bf 0}$, the laminar current is absent.

%In addition, whether the magnetization is present or not, the flow directions of $J^{x}$ and $J^{y}$ are always symmetric and antisymmetric between the two edges, respectively, inheriting the features in  graphene in Fig.~\ref{fig:graphene_spin_current}.

We conclude that for magnetization along principle directions ${\bf S}\parallel\left\{{\hat{\bf x}},{\hat{\bf y}},{\hat{\bf z}}\right\}$, only $\left\{\chi^{xz},\chi^{zx}\right\}$ of the transverse susceptibility are nonzero, which then yield a local spin torque that requires no bias voltage according to Eq.~(\ref{Landau_Lifshitz_dynamics}). However, the transverse spin polarization is alway antisymmetric between the two edges and hence integrates to zero $\sum_{i}d{\bf S}_{i}/dt=0$, indicating no net torque on a macroscopic scale. These features are similar to those in the two-site model in Sec.~\ref{sec:two_site_toy_model}.

\begin{figure*}[htpb]
\begin{center}
\includegraphics[clip=true,width=1.6\columnwidth]{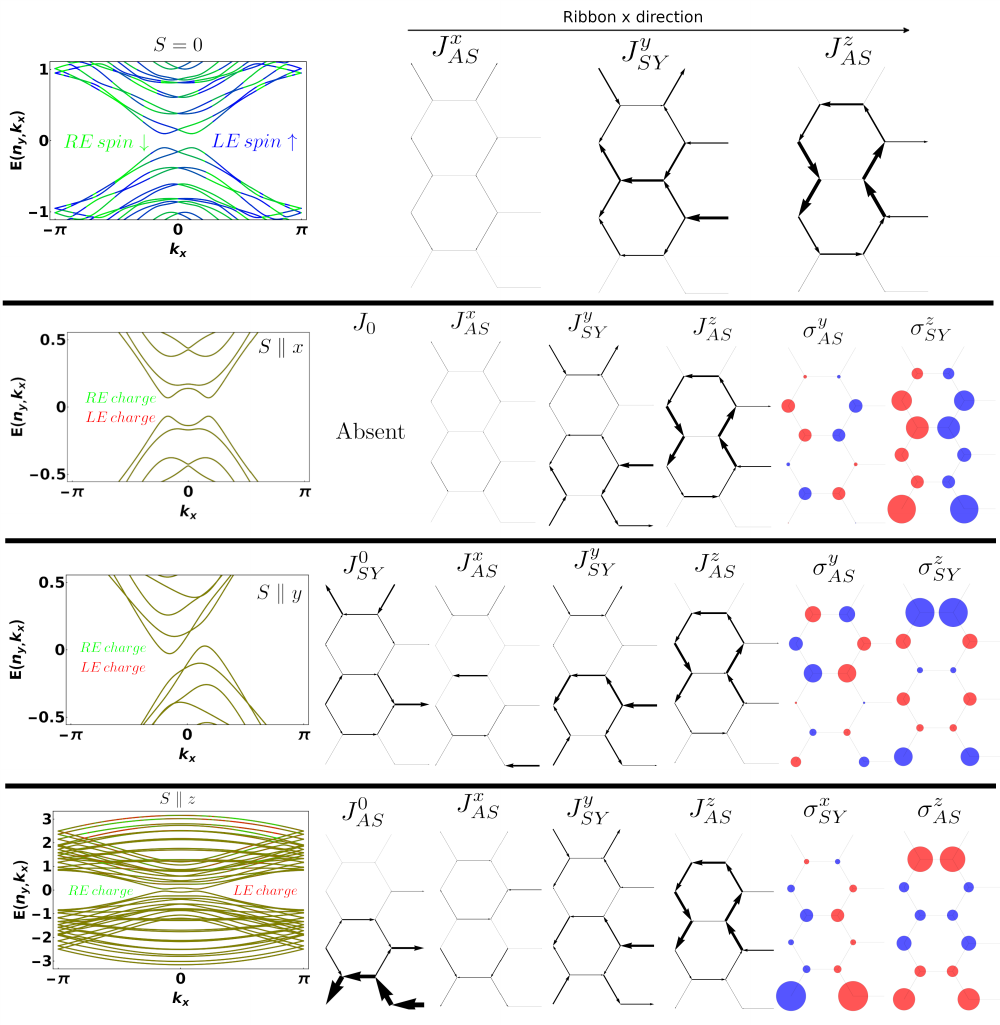}
\caption{Same quantities as in Fig.~\ref{fig:zigzag_ribbon_current} but for glide plane symmetric armchair ribbon. The pattern is symmetric (labeled SY) or antisymmetric (labeled AS) is defined with respect to $P_{\rm arm}^{\rm gli}$ in Eq.~(\ref{armchair_symmetry_operator}). The largest arrow and disk in the figure are of the magnitude $0.024$ and $0.011$, respectively. }
\label{fig:armchair_ribbon_current}
\end{center}
\end{figure*}

\subsection{Persistent currents and spin torques in armchair ribbons \label{sec:persistent_current_armchair}}

The armchair ribbons are simulated by imposing PBC in ${\hat{\bf x}}$ and OBC in ${\hat{\bf y}}$ to Eq.~(\ref{Hamiltonian_graphene_Rashba_mag}). The symmetry between the two edges distinguishes mirror symmetric and the glide plane symmetric armchair ribbons, with the corresponding symmetry operators
\begin{eqnarray}
P_{\rm arm}^{\rm mir}=\left\{P_{y}|\,{\bf 0}\right\},\;\;\;
P_{\rm arm}^{\rm gli}=\left\{P_{y}|\,3/2{\hat{\bf x}}\right\},
\label{armchair_symmetry_operator}
\end{eqnarray}
where $P_{y}$ denotes the mirror operation along ${\hat{\bf y}}$ with respect to the central axis of the ribbon. The symmetry properties of the currents and spin polarization for the mirror and glide plane symmetric armchair ribbons, defined under $P_{\rm arm}^{\rm mir}$ and $P_{\rm arm}^{\rm gli}$, respectively, are identical, so we only present the glide plane symmetric case in Fig.~\ref{fig:armchair_ribbon_current}. The edge spin and charge operators are analogous to those introduced  in Eq.~(\ref{zigzag_edge_charge}) and (\ref{zigzag_edge_spin}), with an exchange of coordinates $x\leftrightarrow y$. The results for different magnetization directions are summarized below.

(i) ${\bf S}={\bf 0}$ : The case $ S=0$ has both the $J^{x}$ and $J^{y}$ caused by RSOC. However, $J^{x}$ is strongly suppressed due to OBC in ${\hat{\bf y}}$ direction. The out-of-plane polarized $J^{z}$ at the edge is clearly visible, and also originates from the counter propagating spins as can be deduced from the color codes of the band structure. These features of spin currents remain valid in the ${\bf S}\neq{\bf 0}$ cases below.

(ii) ${\bf S}\parallel{\hat{\bf x}}$ : For the magnetization pointing along the ribbon, there is no charge current $J^{0}$, and a spin polarization in both transverse directions is induced. The $\sigma^{y}$ component is antisymmetric and, hence, constitutes a non-collinear magnetic order between the two edges. The $\sigma^{z}$ component is symmetric for the two edges, but has an alternating signs along the ribbon, causing non-collinear magnetic order along the ribbon.

(iii) ${\bf S}\parallel{\hat{\bf y}}$ : For magnetization in-plane but perpendicular to the ribbon direction, an asymmetric band structure causes a nonchiral $J^{0}$, and the spin polarization in both transverse directions $\sigma^{x}$ and $\sigma^{z}$ are induced and are asymmetric between the two edges.

(iv) ${\bf S}\parallel{\hat{\bf z}}$ : The out-of-plane magnetization again causes a symmetric band structure. For $+k_{x}$ state that is more localized at the right edge, we find a corresponding $-k_{x}$ state at the left edge, which suggests the existence of a chiral $J^{0}$. Interestingly, this feature is not obvious at low energy, but more prominent for the higher energy states. Comparing this with the ${\bf S}\parallel{\hat{\bf y}}$ result, we again see that the chirality of $J^{0}$ can be controlled by the orientation of the magnetization. The $\sigma^{z}$ component is antisymmetric between the two edges, while $\sigma^{y}$ is symmetric but with sign-alternating along the ribbon. Finally, comparing the spin polarization $\sigma^{\alpha}$ in all three situations ${\bf S}\parallel\left\{{\hat{\bf x}},{\hat{\bf y}},{\hat{\bf z}}\right\}$, we conclude that every component of the local susceptibility tensor $\chi^{\alpha\beta}$ is nonzero, different from that of the zigzag ribbon in Sec.~\ref{sec:persistent_current_zigzag}. Nevertheless, the total torque vanishes in all cases.

%$\sum_{i}d{\bf S}_{i}/dt=0$ due to the antisymmetric $\sigma^{\alpha}$ between the two edges.

%Both transverse spin polarizations are induced, with a symmetric $\sigma^{x}$ and antisymmetric $\sigma^{y}$, and both averages to zero in mesoscopic scale.

%The result again indicates a persistent charge current when the magnetization is perpendicular to the ribbon direction, which is symmetry between the two edges when ${\bf S}\parallel{\hat{\bf y}}$, and antisymmetric when ${\bf S}\parallel{\hat{\bf z}}$. Comparing the band structures at different ${\bf S}$, one also concludes that the asymmetric band structure is not a necessary condition for the charge current. Concerning the spin currents, whether the magnetization ${\bf S}={\bf 0}$ is present or not, an antisymmetric $J^{x}$ (largely suppressed due to OBC in ${\hat{\bf y}}$), symmetric $J^{y}$, and antisymmetric $J^{z}$ near the edge are produced, similar to that in the 2DEG\cite{Reynoso04,Usaj05}.

%In the presence of the magnetization ${\bf S}\neq{\bf 0}$, the symmetries of the spin currents remain. These features are very similar to those in the zigzag ribbon discussed in Sec.~\ref{sec:persistent_current_zigzag}.

%So a magnetization that exists only partially in the system is again required to cultivate this torque.

%\begin{figure}[h]
%\begin{center}
%\includegraphics[clip=true,width=0.95\columnwidth]{armchair_ribbon_spin_polarization}
%\caption{ The spin polarization in the armchair ribbon with different magnetization directions. }
%\label{fig:armchair_ribbon_spin_polarization}
%\end{center}
%\end{figure}

\subsection{Partially magnetized nanoribbons and irregular nanoflakes}

\begin{figure}[h]
\begin{center}
\includegraphics[clip=true,width=0.8\columnwidth]{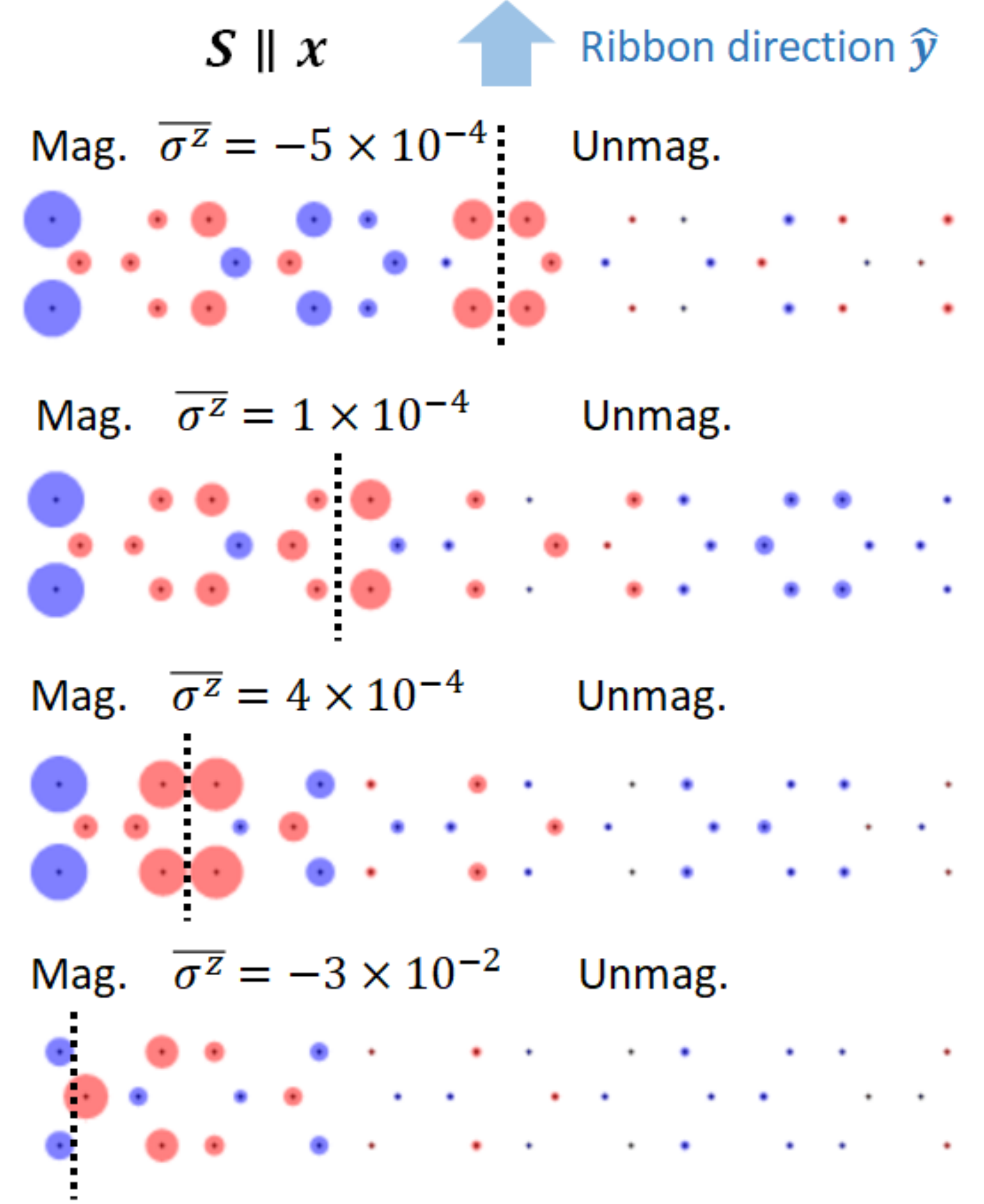}
\caption{The unit cell of partially magnetized zigzag ribbon at magnetization ${\bf S}\parallel{\hat{\bf x}}$, where the transverse spin polarization $\sigma^{z}$ is indicated by the size of disks, with red positive and blue negative. The region to the left of the dotted line is magnetized, and to the right is unmagnetized, and hence the panels from top to bottom represent the magnetization covering $1/2$, $1/3$, $1/6$, and the edge sites of the ribbon. The average spin polarization per site in the magnetized region is indicated by $\overline{\sigma^{z}}$. }
\label{fig:partially_magnetized_zigzag}
\end{center}
\end{figure}

To make the magnetoelectric torque observable, one must overcome the overall cancelling of the torque as seen in our discussion of the nanoribbons. We suggest two situations that the net torque can be nonzero. The first is to make the magnetization spatially varying, e.g. only occupy the region closer to one edge, such that the other edge is idle. Using zigzag ribbon with ${\bf S}\parallel{\hat{\bf x}}$ as an example, in Fig.~\ref{fig:partially_magnetized_zigzag} we indeed see a nonzero net transverse spin polarization in this situation, whose average value per site is of the order of $\overline{\sigma^{z}}\sim 10^{-4}$ when the magnetization only covers a region near the left edge, and it is dramatically enhanced to $\overline{\sigma^{z}}\sim 10^{-2}$ if the magnetization only exists on the edge sites. Following Eq.~(\ref{Landau_Lifshitz_dynamics}), the $\overline{\sigma^{z}}\sim 10^{-4}$ and $J_{ex}\sim 0.1$eV would yield a very large spin torque $d{\bf S}/dt\sim 10$GHz. Although we likely overestimated this torque due to the large $J_{ex}$ and $\lambda_{R}$ in Eq.~(\ref{graphene_parameters}), even if the torque is reduced by two orders of magnitude to $\sim 0.1$GHz, it is still significant.

\begin{figure}[h]
\begin{center}
\includegraphics[clip=true,width=0.95\columnwidth]{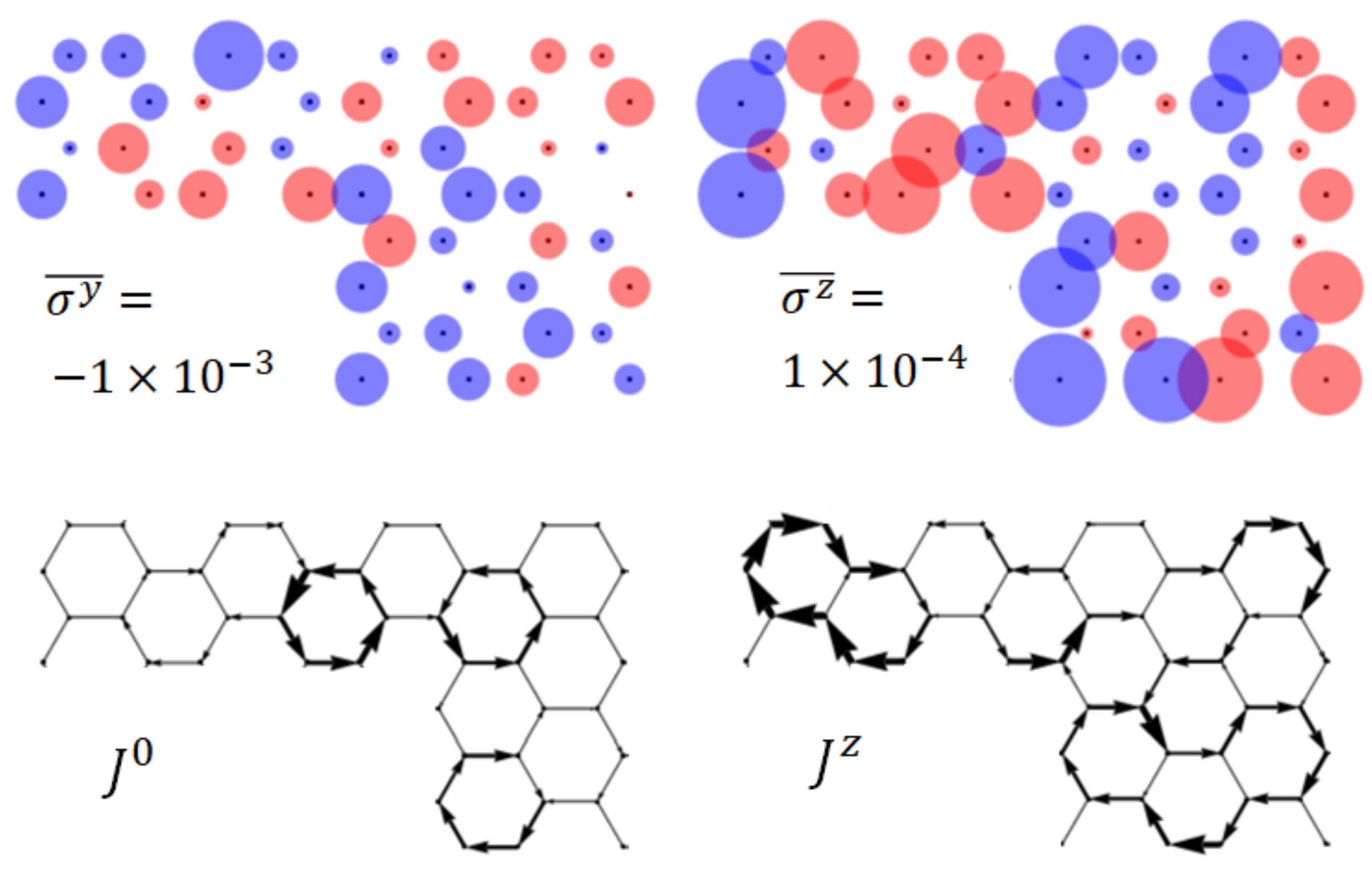}
\caption{An L-shape nanoflake with magnetization ${\bf S}\parallel{\hat{\bf x}}$, where we show the transverse spin polarizations $\sigma^{y}$ and $\sigma^{z}$, charge current $J^{0}$, and out-of-plane polarized spin current $J^{z}$. The $\overline{\sigma^{y,z}}$ represents the average spin polarization per site. }
\label{fig:nanoflake_irregular}
\end{center}
\end{figure}

The second proposal is to use graphene nanoflakes with an irregular shapes, which may help to generate nonuniform distribution of spin polarization that does not sum to zero\cite{Ma11,Weymann12,Guclu13,Luo14,Szalowski15}, given that the two edges are not equivalent or it may even be ambiguous to identify two opposite edges. In the example shown in Fig.~\ref{fig:nanoflake_irregular}, we see that the L-shape nanoflake indeed yields a net $\overline{\sigma^{y}}$ and $\overline{\sigma^{z}}$ per site of the order of $10^{-3}$ tp $10^{-4}$ when magnetization points at ${\bf S}\parallel{\hat{\bf x}}$, indicating a net torque of significant strength. In addition, despite the open boundary in all directions, still equilibrium currents $\left\{J^{0},J^{x},J^{y},J^{z}\right\}$ exist. These currents turn into networks of local currents, suggesting that they survive even in realistic experimental situations of small and open boundary nanoflakes.

\section{Conclusions \label{sec:conclusions}}

In summary, we demonstrate that graphene nanoribbons due to RSOC and geometric confinement display a peculiar magnetic response, including a bias voltage-free spin torque, chiral and nonchiral edge currents, and equilibrium spin currents. Using a two-site toy model, we could analytically show features such as a transverse susceptibility caused by geometric confinement and linear in the RSOC strength. Numerical calculation suggests that the same features occur in nanoribbons. Given the controllability of RSOC by gate voltage\cite{Yang16,Wang16_3,Yang17,Safeer19,Ghiasi19,Benitez20}, the transverse susceptibility can serve as a gate-voltage-induced magnetoelectric torque distinct from the usual current-induced spin-orbit torque. In a zigzag ribbon, this torque tends to create noncollinear spin polarization between the two edges, and in armchair ribbon the noncollinear order is not only between the edges but can also be along the ribbon. Although the net torque sums to zero in uniformly magnetized ribbons, it can be nonzero if the magnetization only covers parts of the ribbon, say, closer to one edge, or in nanoflakes of irregular shapes. We estimate a net torque that can reach sub-GHz magnitude, pointing to the possibility of practical applications.

We further confirm that the equilibrium edge charge current $J^{0}$ predicted for 2DEG, when the magnetization points perpendicular to the edge, also exists in Rashba nanoribbons. Moreover, we discovered that $J^{0}$ can be nonchiral due to the asymmetric band structure, or chiral due to counter propagating eigenstates localized at opposite edges, depending on the magnetization pointing in-plane or out-of-plane. In addition, the in-plane polarized persistent spin currents $J^{x}$ and $J^{y}$ that exist ubiquitously in 2D Rashba systems also manifest in nanoribbons, but the component flowing in the confined direction is strongly suppressed. Besides, the RSOC causes an out-of-plane polarized chiral edge spin current $J^{z}$ even in the absence of magnetization. This discovery poses a challenge to distinguish $J^{z}$ from that of purely topological origin in graphene-based topological insulators containing RSOC, such as the Kane-Mele model\cite{Kane05_2,Kane05}. Generally, the energy gap does not affect the existence of these currents and spin torques, since they are contributed from all the eigenstates in the Fermi sea, not only the low energy flat band edge states that in some cases become chiral. In addition, all these currents survive even in nanoflakes that have open boundary in every direction. We anticipate that the controllability of RSOC by gate voltage and the magnetization by magnetic field offers a practical way to engineer these effects, which may help to realize them for practical purposes, such as building graphene-based spintronic devices\cite{Pesin12_2,Han14,Roche15}.

\section{Acknowledgement}

We thank exclusively J. C. Egues and A. Zegarra for fruitful discussions. This work is financially supported by the productivity in research fellowship from CNPq.

\appendix

\section{Explicit form of the charge and spin current operators \label{apx:charge_spin_current_operator}}

To calculate the commutators in Eq.~(\ref{charge_spin_continuity_eq}), we observe that the hopping and the Rashba Hamiltonians take the following general form
\begin{eqnarray}
H_{t}+H_{R}=\sum_{i\in A}\sum_{\eta}\sum_{\alpha\beta}\left\{T_{\alpha\beta}^{\eta}c_{i\alpha}^{\dag}c_{i+\eta\beta}
+T_{\alpha\beta}^{\eta\ast}c_{i+\eta\beta}^{\dag}c_{i\alpha}\right\},
\nonumber \\
\end{eqnarray}
where $T_{\alpha\beta}^{\eta}$ is the hopping amplitude that the spin $\beta$ at site $i+\eta$ becomes spin $\alpha$ at site $i$, and $T_{\alpha\beta}^{\eta\ast}$ is the complex conjugate of this amplitude. The results of the commutators give
\begin{eqnarray}
J_{i,i+\eta}^{0}&=&\frac{ia}{\hbar}\sum_{\alpha\beta}\left\{T_{\alpha\beta}^{\eta}c_{i\alpha}^{\dag}c_{i+\eta\beta}
-T_{\alpha\beta}^{\eta\ast}c_{i+\eta\beta}^{\dag}c_{i\alpha}\right\},
\nonumber \\
J_{i,i+\eta}^{a}&=&\frac{ia}{\hbar}\sum_{\alpha\beta}\sum_{\mu\nu}
\left\{T_{\alpha\beta}^{\eta}\delta_{\alpha\nu}\sigma_{\mu\nu}^{a}c_{i\mu}^{\dag}c_{i+\eta\beta}\right.
\nonumber \\
&&\left.-T_{\alpha\beta}^{\eta\ast}\delta_{\mu\alpha}\sigma_{\mu\nu}^{a}c_{i+\eta\beta}^{\dag}c_{i\nu}\right\}.
\end{eqnarray}
For ${\boldsymbol\eta}=\left(-{\boldsymbol\delta_{1}},-{\boldsymbol\delta_{2}},-{\boldsymbol\delta_{3}}\right)$, the complete list of nonzero hopping amplitudes are
\begin{eqnarray}
&&T_{\uparrow\uparrow}^{-\delta_{i}}=T_{\downarrow\downarrow}^{-\delta_{i}}=t\;\;\;{\rm for}\;i=\left\{1,2,3\right\},
\nonumber
\end{eqnarray}
\begin{eqnarray}
&&T_{\uparrow\downarrow}^{-\delta_{1}}=i\lambda_{R}\left(-\frac{\sqrt{3}}{2}-\frac{i}{2}\right),\;\;\;
T_{\downarrow\uparrow}^{-\delta_{1}}=i\lambda_{R}\left(-\frac{\sqrt{3}}{2}+\frac{i}{2}\right),
\nonumber
\end{eqnarray}
\begin{eqnarray}
&&T_{\uparrow\downarrow}^{-\delta_{2}}=i\lambda_{R}\left(\frac{\sqrt{3}}{2}-\frac{i}{2}\right),
\;\;\;
T_{\downarrow\uparrow}^{-\delta_{2}}=i\lambda_{R}\left(\frac{\sqrt{3}}{2}+\frac{i}{2}\right),
\nonumber
\end{eqnarray}
\begin{eqnarray}
&&T_{\uparrow\downarrow}^{-\delta_{3}}=-\lambda_{R},\;\;\;
T_{\downarrow\uparrow}^{-\delta_{3}}=\lambda_{R},
\end{eqnarray}
and $T_{\alpha\beta}^{\delta_{i}}=-T_{\alpha\beta}^{-\delta_{i}}$ for ${\boldsymbol\eta}=\left({\boldsymbol\delta_{1}},{\boldsymbol\delta_{2}},{\boldsymbol\delta_{3}}\right)$. The explicit forms of the current operators are then given by
\begin{eqnarray}
&&J_{i,i+\eta}^{0}=\frac{ia}{\hbar}\left\{t\,c_{i\uparrow}^{\dag}c_{i+\eta\uparrow}
-t\,c_{i+\eta\uparrow}^{\dag}c_{i\uparrow}+t\,c_{i\downarrow}^{\dag}c_{i+\eta\downarrow}
-t\,c_{i+\eta\downarrow}^{\dag}c_{i\downarrow}\right.
\nonumber \\
&&\left.+T_{\uparrow\downarrow}^{\eta}\,c_{i\uparrow}^{\dag}c_{i+\eta\downarrow}
-T_{\uparrow\downarrow}^{\eta\ast}\,c_{i+\eta\downarrow}^{\dag}c_{i\uparrow}
+T_{\downarrow\uparrow}^{\eta}\,c_{i\downarrow}^{\dag}c_{i+\eta\uparrow}
-T_{\downarrow\uparrow}^{\eta\ast}\,c_{i+\eta\uparrow}^{\dag}c_{i\downarrow}
\right\},
\nonumber
\end{eqnarray}
\begin{eqnarray}
&&J_{i,i+\eta}^{x}=\frac{ia}{\hbar}\left\{t\,c_{i\downarrow}^{\dag}c_{i+\eta\uparrow}
-t\,c_{i+\eta\uparrow}^{\dag}c_{i\downarrow}+t\,c_{i\uparrow}^{\dag}c_{i+\eta\downarrow}
-t\,c_{i+\eta\downarrow}^{\dag}c_{i\uparrow}\right.
\nonumber \\
&&\left.+T_{\uparrow\downarrow}^{\eta}\,c_{i\downarrow}^{\dag}c_{i+\eta\downarrow}
-T_{\uparrow\downarrow}^{\eta\ast}\,c_{i+\eta\downarrow}^{\dag}c_{i\downarrow}
+T_{\downarrow\uparrow}^{\eta}\,c_{i\uparrow}^{\dag}c_{i+\eta\uparrow}
-T_{\downarrow\uparrow}^{\eta\ast}\,c_{i+\eta\uparrow}^{\dag}c_{i\uparrow}
\right\},
\nonumber
\end{eqnarray}
\begin{eqnarray}
&&J_{i,i+\eta}^{y}=\frac{a}{\hbar}\left\{-t\,c_{i\downarrow}^{\dag}c_{i+\eta\uparrow}
-t\,c_{i+\eta\uparrow}^{\dag}c_{i\downarrow}+t\,c_{i\uparrow}^{\dag}c_{i+\eta\downarrow}
+t\,c_{i+\eta\downarrow}^{\dag}c_{i\uparrow}\right.
\nonumber \\
&&\left.-T_{\uparrow\downarrow}^{\eta}\,c_{i\downarrow}^{\dag}c_{i+\eta\downarrow}
-T_{\uparrow\downarrow}^{\eta\ast}\,c_{i+\eta\downarrow}^{\dag}c_{i\downarrow}
+T_{\downarrow\uparrow}^{\eta}\,c_{i\uparrow}^{\dag}c_{i+\eta\uparrow}
+T_{\downarrow\uparrow}^{\eta\ast}\,c_{i+\eta\uparrow}^{\dag}c_{i\uparrow}
\right\},
\nonumber
\end{eqnarray}
\begin{eqnarray}
&&J_{i,i+\eta}^{z}=\frac{ia}{\hbar}\left\{t\,c_{i\downarrow}^{\dag}c_{i+\eta\uparrow}
-t\,c_{i+\eta\uparrow}^{\dag}c_{i\downarrow}-t\,c_{i\uparrow}^{\dag}c_{i+\eta\downarrow}
+t\,c_{i+\eta\downarrow}^{\dag}c_{i\uparrow}\right.
\nonumber \\
&&\left.+T_{\uparrow\downarrow}^{\eta}\,c_{i\uparrow}^{\dag}c_{i+\eta\downarrow}
-T_{\uparrow\downarrow}^{\eta\ast}\,c_{i+\eta\downarrow}^{\dag}c_{i\uparrow}
-T_{\downarrow\uparrow}^{\eta}\,c_{i\downarrow}^{\dag}c_{i+\eta\uparrow}
+T_{\downarrow\uparrow}^{\eta\ast}\,c_{i+\eta\uparrow}^{\dag}c_{i\downarrow}
\right\}.
\nonumber \\
\end{eqnarray}
We then evaluate the equilibrium charge and spin current numerically by their expectation values
\begin{eqnarray}
\langle J_{i,i+\eta}^{a}\rangle=\sum_{n}\langle n|J_{i,i+\eta}^{a}|n\rangle\,f(E_{n}),
\end{eqnarray}
where $|n\rangle$ is the eigenstate of the lattice Hamiltonian, and $f(E_{n})=\left(e^{E_{n}/k_{B}T}+1\right)^{-1}$ is the Fermi distribution function. Note that this strategy includes all the eigenstates in the Fermi sea, in contrast to previous theories that consider only the low energy states\cite{Zhang14,Li17}. Often times we ignore that bracket $\langle{\hat{\cal O}}\rangle\equiv{\hat{\cal O}}$ of the expectation value for convenience.

\bibliography{Literatur}

\end{document}